\documentclass[aps,twocolumn,prl]{revtex4}
\usepackage{graphicx}
\usepackage{amssymb}

\begin{document}

%\draft

%\wideabs{
\title{Sympathetic Cooling with Two Atomic Species in an Optical Trap}

\author{M. Mudrich}
\author{S. Kraft}
\author{K. Singer}
%\affiliation{Max-Planck-Institut f\"ur Kernphysik, Postfach 103980, 69029 Heidelberg,
%Germany}
\author{R. Grimm}
\altaffiliation{
 Permanent address: Institut f\"{u}r Experimentalphysik, Universit\"at
Innsbruck, 6020 Innsbruck, Austria}
%\affiliation{Max-Planck-Institut f\"ur Kernphysik, Postfach 103980, 69029 Heidelberg,
%Germany}
\author{A. Mosk}
%\affiliation{Max-Planck-Institut f\"ur Kernphysik, Postfach 103980, 69029 Heidelberg,
%Germany}
\author{M. Weidem\"{u}ller}
\affiliation{Max-Planck-Institut f\"ur Kernphysik, Postfach 103980, 69029 Heidelberg,
Germany\\
http://www.mpi-hd.mpg.de/ato/lasercool}

%}
\date{}
\begin{abstract}
We simultaneously trap ultracold lithium and cesium atoms in an
optical dipole trap formed by the focus of a CO$_2$ laser and study the exchange of thermal energy
between the gases.
The
cesium gas, which is optically cooled to $20\; \mu$K,
 efficiently decreases the temperature of the lithium gas
 through sympathetic cooling. The measured cross section for thermalizing
 $^{133}$Cs-$^7$Li
 collisions is
$8 \times 10^{-12}$ cm$^2$, for both species in their lowest hyperfine ground state.
Besides thermalization, we observe evaporation of lithium purely
through elastic cesium-lithium collisions (sympathetic
evaporation).
\end{abstract}

\maketitle
Exchange of heat through thermal contact between two ensembles
is the most obvious and ubiquitous thermodynamic process.
 In the context of cold gases, this process can be utilized
to sympathetically cool one component in a mixture. It is particularly useful
  if the other component can be cooled
 efficiently in a direct way, e.g. by optical methods.
 Sympathetic cooling has
 been
demonstrated with charged particles which thermalize through the long-range Coulomb
 interaction~\cite{Larson1986}.
For neutral atoms and molecules one can use
helium gas as a cooling agent which can be cooled cryogenically
   to well below $1$ K~\cite{deCarvalho1999a}.
 Only recently,
sympathetic cooling has been successfully extended to the ultracold
regime~\cite{Myatt1997a,Truscott2001a,Schreck2001a,Esslinger2001x,Modugno2001}.
 Quantum degeneracy of
bosons and fermions was obtained by thermalization between atoms of the same species in
 different internal
states~\cite{Myatt1997a}, between two isotopes of
 the same species~\cite{Truscott2001a,Schreck2001a} and,
 as a recent highlight, between
atoms of different species~\cite{Modugno2001}.
The degree of quantum degeneracy obtained in those experiments was
finally limited by two important loss processes: exoergic
collisions between magnetically trapped atoms, and evaporation,
which decreases the heat capacity of the cooling agent.

In this Letter, we explore the thermodynamics of sympathetic cooling
for a mixture of two
different atomic gases at ultralow temperatures. We present an intrinsically loss-free
 approach which can be
generalized to a vast range of atomic species and even molecules.
Both the coolant gas and the gas of interest are in their internal
energetic ground state, and
confined to the same volume by the optical dipole force~\cite{Grimm00}.
%Loss of particles through exoergic binary collisions is
%excluded since the particles can be stored in their internal energetic ground state.
The coolant gas is cooled by
pure optical techniques~\cite{Boiron1998a} which can be applied in an optical dipole trap without
 significant loss.

 Our
model system is a mixture of cesium and lithium in an extremely far-detuned optical dipole trap.
 Cesium is an
excellent cooling agent, since it can be optically cooled to very low temperatures,
 mainly due to its high mass.
Lithium, on the other side, is difficult to cool optically.
In addition, it is nearly an ideal gas due the
very small cross section for elastic Li-Li collisions~\cite{Schreck2001a}.
Thermalization thus occurs purely
through collisions with the cesium atoms.
As a second important thermodynamic process besides thermalization,
 we study evaporation through
Li-Cs collisions (sympathetic evaporation)
 in traps with finite depth.
Both processes allow an independent measurement of the a priori
unknown interspecies scattering cross-section.

To trap a mixture of ultracold species, the focus
 of a CO$_2$ laser beam, at $\lambda=10.6\; \mu$m, constitutes an almost perfect
realization of a conservative trapping potential
(quasi-electrostatic trap, QUEST)~\cite{Grimm00,takekoshi96,O'Hara1999c,Engler00}.
 Atoms and molecules can be stored in any
internal state, especially the energetic ground
state. For atoms in
the lowest-energy hyperfine state, energy-releasing processes in
two-body collisions cannot occur.
Due to the large detuning of the laser frequency from
any atomic resonances, heating through photon scattering can be
completely neglected.

Using a commercial sealed-tube laser, we obtain a power of 108 W in
 the vacuum chamber, focused to a waist
$w_0=86\;\mu$m.
 For a Gaussian laser beam, with power $P$ and waist $w_0$
 the trap depth is given by $U_0= \alpha_{\rm stat} P/(\pi \varepsilon_0 c
 w_0^2)$, which yields $U_0^{\rm Cs}/k_B=0.85$ mK and $U_0^{\rm Li}/k_B=0.34$ mK.
 By measuring the oscillation frequencies of Cs atoms in the trap,
 $\omega_{\rm x,y}/2 \pi=850 $ Hz and $\omega_{z}/2 \pi = 18$ Hz, we
find our trap is slightly deformed axially,
probably due to stable  multimode operation of the laser
\cite{MoskAppPB2001}.

Atoms are transferred into the dipole trap from
magneto-optical traps (MOT) for lithium and cesium which are
superimposed on the focus of the CO$_2$ laser beam. Both MOTs are
loaded from Zeeman-slowed atomic beams.
The lifetime of the gas is limited only by
rest-gas collisions to $\approx 100$ s.
Details of our apparatus and of the
procedure for loading atoms into the QUEST are given in Refs.~\cite{MoskAppPB2001,Engler00}.

Transferred separately, we obtain $5\times 10^5$ Cs atoms in the QUEST,
which we cool by polarization gradient cooling to $20\; \mu$K.
We find that the temperature is lower for a shallower trap, while
in free space we reach temperatures below $3 \;\mu$K. Apparently
the Stark shifts caused by the QUEST influence the
cooling process in a way which is not yet understood.
The Cs is optically pumped into the
lowest hyperfine state ($F=3$), and since we compensate the magnetic field,
the atoms are distributed evenly over the degenerate $m_F$ states.
The peak density of Cs atoms in the QUEST is  $\sim 2 \times 10^{11}$
cm$^{-3}$. Since the scattering length of the Cs is larger than
the typical relative wavevector in collisions, the internal
thermalization of the Cs gas is dominated by
 the unitary maximum $s$-wave scattering cross-section~\cite{Arndt1997a}
 $\sigma_\mathrm{CsCs} = 4 \pi/k^2$, where $k$
 is the relative wavevector. We use the expression for distinguishable atoms
  since the majority of the collisions
 are between atoms in different $m_F$ states. The collision rate is
 approximately 7 per second, and thermalization
 through Cs-Cs collisions is estimated to take $\sim 2$ seconds~\cite{Arndt1997a}.

The situation for Li is different: since the temperature
reached in the MOT is higher than the trap depth, only
approximately $10^5$ atoms are transferred. Numerical
simulations indicate these atoms occupy a truncated thermal distribution
with an internal energy of $ \sim 3 k_B\times\, 75\; \mu$K,
which only very weakly depends on the MOT temperature.
The lithium atoms are optically pumped into
the $F=1$ hyperfine state, and in the absence of a magnetic guiding field
distribute themselves
over the degenerate $m_F$ states. As the scattering
length of Li in the $F=1$ state is anomalously small~\cite{Schreck2001a},
 the distribution does not thermalize
on the 100 s timescale of our experiment.

In order to simultaneously trap the two species, a
Cs MOT is loaded and the atoms transferred into the QUEST in optical molasses.
Subsequently, a
Li MOT is loaded at a slightly different position, to minimize
light-induced atom loss in a two-species MOT~\cite{schloeder99}. Once it reaches the desired
density, the Li MOT is compressed and overlapped with the QUEST
for a few ms to optimize transfer of Li atoms. We can
simultaneously trap typically $4 \times 10^4$ Li atoms with up to
$10^5$
Cs.

After a variable interaction time in the QUEST, usually a few seconds,
we analyze the gas sample. The
temperature and atom number of Cs are determined by absorption
imaging and time-of-flight velocimetry~\cite{Ketterle1999a}.
However, the lithium gas in our trap is too dilute to use
absorption imaging.
We determine the temperature of the lithium gas using
a release- recapture method.
The dipole trap potential is suddenly turned off for a variable
 ballistic expansion time of order 1 ms.
After this time, the dipole trap is turned on again for $\sim 500$ ms to recapture
the atoms that are still close to the center of the trap, while
the others are lost. The number of remaining atoms can then be measured by the
 fluorescence of a MOT.

\begin{figure}
\center
\includegraphics[width=8 cm]{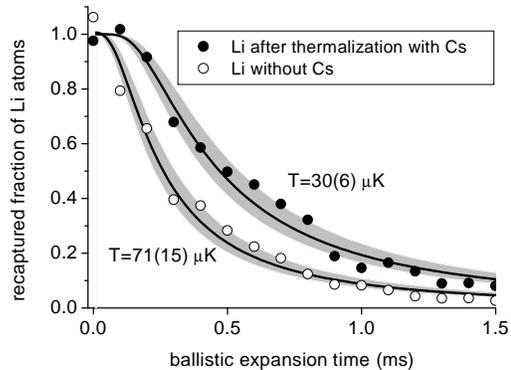}
\caption{Release-recapture thermometry for Li atoms. The fraction of atoms that were recaptured in the dipole trap
is shown as a function of the time the trap was turned off.
 Open circles: Li without Cs.
 Closed circles: Li thermalized with Cs for 5 s.
Solid line: model fits to Eq.~(\ref{Eq:gaussianapproximation})
The gray area denotes the 20\% uncertainty in the temperature.
   }
\label{fig:RRcurve}       % Give a unique label
\end{figure}

Figure \ref{fig:RRcurve}  shows two such
release-recapture measurements: one for a pure Li ensemble and one for a Li ensemble that has
 been trapped simultaneously with cold Cs.
 The
pure Li gas is seen to
be lost from the trap in $\sim 0.4$ ms, while the ensemble of Li that has
been in thermal contact with cold Cs ($T_{\rm Cs}= 28\; \mu$K initially)
  is seen to contain many more slow atoms, which
 indicates cooling.
  To quantitatively understand this cooling process
  we compare the release-recapture measurements to
calculations.

For a gas at low temperature ($k_B T \ll U_0$) in an extremely
elongated Gaussian trap $(\omega_{\rm ax} \ll \omega_{\rm rad})$,
we derive an analytical
approximation of the number of recaptured atoms,
\begin{equation}
\label{Eq:gaussianapproximation}
N_{\rm r}(t)=\frac{N_0}{1-e^{-\eta}}
\left[
1-\exp\left(- \eta W(\tilde{t}^2) / \tilde{t}^{2}  \right)
\right].
\end{equation}
Here $w_0$ is the Gaussian beam waist, $\eta=U_0/k_B T$, $\tilde{t}^2= m w_0^2 t^2 /4U_0$, and  $W(z)$ is the product logarithm function, satisfying $W(z)
\exp(W(z))=z$. The analytical approximation of Eq.~(\ref{Eq:gaussianapproximation}) agrees with
numerical simulation
 for temperatures up to $k_B T \sim 0.15 \; U_0$.
Full numerical calculation of $N_{\rm r}(t)$ is possible for any $T$ and shows that
 for $k_B T>0.2\; U_0$  ($75$ mK for Li)
 the width of the recapture curve becomes essentially independent of temperature, i.e.,
release-recapture measurements no longer provide reliable temperature data.
Hence  Eq.~(\ref{Eq:gaussianapproximation}) can be used
over all of the useful parameter range of this thermometry
method. For the temperature range of interest, we estimate the
accuracy of this method to be better than $20\%$.

The release-recapture curve of the pure Li in Fig.~\ref{fig:RRcurve}
can be reproduced by assuming a thermal distribution at $T_{\rm Li}=71\pm 15\,\mu$K.
The sympathetically cooled Li ensemble is well-described by  $T_{\rm Li}= 30 \pm
6\,\mu$K, which is equal to the Cs temperature within
the uncertainty margins, indicating that the gases have fully
thermalized.

\begin{figure}
\center
\includegraphics[width=8 cm]{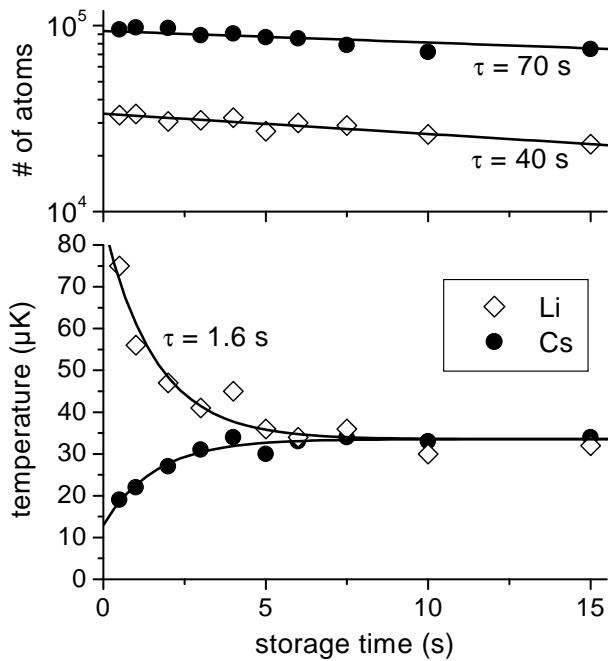}
\caption{ Evolution of Cs and Li temperatures (lower graph) and atom numbers (upper graph)
 during simultaneous storage.
Line: exponential fit, thermalization time
$1.6(2)$ s.}
\label{fig:Therm}       % Give a unique label
\end{figure}

We measure the thermalization time by making a series of
temperature measurements for both components, as seen in
Fig.~\ref{fig:Therm}. The Cs and Li temperatures converge to the
same value $\bar{T}=33 \, \mu K$ to within $2\%$. The measured atom
numbers show that no significant loss of either Li or Cs occurs
during thermalization.
%%%%%%%%%%%%%%%%%%%%%%%%%%%%%%%%%%%%%%%%%%%%%%%%%%%%%%%%%%%%%%%%%%%%%
The thermalization time is $1.6(2)$ s, indeed much shorter than the rest-gas induced decay time
of the trapped gas. In a gas mixture with energy independent
 cross-section, the average energy transferred per collision is
 $k_B \Delta T \xi$, where $\Delta T$ is the temperature difference
 between the components, and $\xi=4 m_1 m_2/(m_1+m_2)^2$ is the
 reduction factor due to the mass difference
 ($\xi=0.19$ for a Li-Cs mixture).
  The heat capacity of a harmonically
 trapped atomic gas is $3 k_B$ per atom, leading to
an average number of $3/\xi$ collisions per Li atom (assuming $N_{\rm Cs} \gg N_{\rm Li}$)
 needed for
thermalization
  \cite{MoskAppPB2001,Delannoy2001,Arndt1997a}.
The thermalization rate is related to the collision rate per Li
atom $\gamma_{\rm coll}$  by
\begin{equation}
\gamma_{\rm therm}=-\frac{1}{\Delta T}\frac{d}{dt} \Delta T =
\frac{\xi}{3} \frac{N_{\rm Cs}+N_{\rm Li}}{N_{\rm Cs}} \gamma_{\rm
coll}
\end{equation}
From the measured thermalization rate we calculate
$\gamma_{\rm coll}= 7(1)$~s$^{-1}$.

 Surprisingly, the Li atoms
have as high a collision rate with Cs as the Cs atoms themselves, even
though the Cs-Cs collisions have the maximum $s$-wave
cross-section.
 The explanation for this lies in the small unitary
cross section and the low thermal velocity for Cs, both a result of
the high mass. The collision-enhancing effects of the low Li mass
 compensate the loss
of thermalization efficiency that is caused by the mass ratio factor $\xi$.
The inferred cross-section for Li-Cs collisions is $\sigma_{\rm
LiCs}= 0.8(4)
\times 10^{-11}$ cm$^2$, where the main limitation of the accuracy is the
determination of the absolute densities. Assuming $s$-wave scattering dominates,
we find an $s$-wave scattering length
$|a_{\rm LiCs}|=(\sigma_{\rm LiCs}/4\pi)^{1/2}= 180^{+40}_{-50}
\,a_0$. This value is smaller than the inverse of the  wavevector $k$ at thermal
energies, but for superthermal collisions, such as play a role in
evaporation, one would have to take into account the energy dependence of
the $s$-wave cross section~\cite{Arndt1997a}. For thermalizing collisions we estimate
 only a minor influence in our case. Adequate theory
for thermalization and evaporation with energy-dependent
cross-sections is still lacking.
%\begin{equation}
%\label{eq:sigmak}
%\sigma_{LiCs}(k)=4 \pi \frac{|a_{\rm LiCs}|^2}{1 + k^2|a_{\rm
%LiCs}|^2}.
%\end{equation}

In a trap of finite depth, thermalizing collisions lead to
evaporation of particles, as atoms are scattered into untrapped states in the
high-energy tail of the thermal distribution. This evaporation is
practically absent in two cases: if the temperature is much lower
than the trap depth (as is the case for pure Cs samples in our trap) or if
the collision rate is extremely small (as is the case for pure Li samples in
our trap). The thermalizing Li-Cs collisions enable, besides cooling, evaporation of
the Li from the trap.  The Li evaporates even under
 conditions where the Cs evaporation is negligible for two reasons:
The Li trap depth is smaller and the collision rate for Li atoms
exceeds that for Cs atoms. Remarkably, since a Li atom needs
energy to evaporate from the trap, this sympathetic evaporation of the Li
represents transfer of thermal energy from a {\em cold} (Cs) to a {\em hot}
(Li)
ensemble.

\begin{figure}
\center
\includegraphics[width=8 cm]{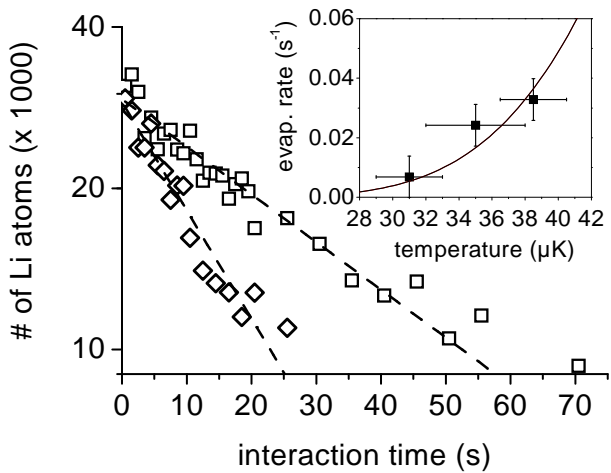}
\caption{Evaporation of Li atoms through Li-Cs collisions. Main graph:
Number of Li atom vs. interaction time, $\diamondsuit:\;T=38 \mu$K; $\square:\;T=30 \mu$K.
Inset: Evaporation loss rate (corrected for rest-gas losses) vs.
temperature. Solid line: Model, see text.}
\label{fig:evap}       % Give a unique label
\end{figure}

The evaporation probability per collision
in an ensemble where the Cs and Li are thermalized is,
 assuming energy-independent cross-sections,
 $P_{\rm evap} \sim \eta \exp(-\eta)$,
where $\eta=U_0^{\rm
Li}/k_BT$~\cite{MoskAppPB2001}.
In numerical simulations for the relevant mass ratio and $\eta$
we  find an atom loss rate
\begin{equation}
\gamma_{\rm evap} \simeq 0.5 \; \eta e^{-\eta} \gamma_{\rm coll}.
\label{eq:tauevap}
\end{equation}
In particular, the simulations confirm that $P_{\rm evap}$
is almost independent of the mass ratio
as long as $\eta > 2/ \xi$.
In Fig.~\ref{fig:evap} the evolution of the number of trapped Li atoms in
thermal contact with Cs is shown for different temperatures. The
inset shows $\gamma_{\rm evap}$ as a function of temperature.
We fit the model of Eq.~(\ref{eq:tauevap}) to the
data, with $\gamma_{\rm coll}$ as  the only free parameter, and obtain a fit at  $\gamma_{\rm
coll} \sim 50$ s$^{-1}$, which results in an effective
cross-section for evaporation of about $5 \times 10^{-11}$ cm$^{2}$,
with an estimated uncertainty of about a factor $3$. This
is of the same order of magnitude as the thermalization
cross-section, but  contributions due to energy-dependence of the
cross-section (e.g., the onset of $p$-wave scattering) cannot be excluded.

Sympathetic evaporation poses a limit to thermalization in shallow traps:
a considerable fraction of the atoms is lost during the first thermalization time
for traps with
$U_0/k_B\bar{T}< W(3 /\xi)$
($\bar{T}$ is the average temperature, $W$ is the product logarithm function). For the  Li-Cs mass ratio,
thermalization without significant loss is possible if $\eta > 4$.
In the initial few collisions, the Li distribution
is far from thermal equilibrium with the Cs, and the evaporation loss cannot be
estimated analytically. In numerical
simulations we see a small loss of Li atoms in the initial few
collisions, which is $< 10 \%$ if $T_{\rm Cs}/U_0^{\rm
Li}<0.1$, in qualitative agreement with the experimental data in
fig.\ref{fig:evap}.

%%%%%%%%%%%%%%%%%%%%%%%%%%%%%%%%%%%%%%%%%%%%%%%%%%%%%%%%%%%%%%%%%%%%%

The usefulness of sympathetic cooling depends strongly on the
timescale of thermalization. To reach the quantum degenerate
regime in our trap geometry, the temperature of the Li gas must be decreased by
two orders of magnitude from the initial temperature after loading, which takes $\sim 5$
thermalization times, assuming the Cs is optically cooled to be
always much below the Li temperature.
This means only
an average number of $\sim 50$ Li-Cs collisions  are needed for
sympathetic cooling, compared with more than 1000 Li-Li
collisions for evaporative cooling of Li to BEC.

The heat which the Li transfers to the Cs can be removed by
repeated pulsed optical cooling. This is more efficient than
continuous cooling since the light-induced inelastic collisions
can act only during the short ($20$ ms) cooling pulses, while
thermalization continues over longer time. In this way, we have
been able to reach Li temperatures down to 25 $\mu$K. The main
limitation for reaching lower temperatures is the base temperature
of the Cs polarization-gradient cooling, which appears to be limited by the trapping potential
and the high Cs
density. Different optical cooling
schemes, or operation at lower Cs density, may provide lower Cs temperatures and ultimately an
evaporation-free route to quantum degeneracy of both the fermionic
and bosonic lithium isotopes.

Another intriguing aspect of the Li-Cs mixture is the formation of cold heteronuclear molecules.
By loss-free
sympathetic cooling, we reach densities of $2 \times 10^{10}$ cm$^{-3}$ and $2 \times 10^{11}$
 cm$^{-3}$ for lithium and cesium, respectively, which provide
good starting conditions for photoassociation experiments~\cite{Masnou2001}.
Once the cold molecules are formed they may be
stored in the quasi-electrostatic trap together with the atomic gases as a first step towards cold quantum
chemistry. In addition, the large electric dipole moment of the LiCs dimer can be used to manipulate and control
internal and external degrees of freedom of the molecules.

{\acknowledgements
 We are indebted to D. Schwalm for
generous support in many respects, and we acknowledge stimulating
discussions with H. A. Weidenm\"uller and A. N. Salgueiro.
The work of A.M. is supported by a Marie-Curie fellowship from the
European Community programme IHP
under contract number CT-1999-00316. The project is supported by the
Deutsche Forschungsgemeinschaft (WE 2661/1-1).
}

\end{document}